\documentclass[11pt]{article}

\usepackage{amsmath,amssymb}
\usepackage{graphicx}
\newcommand{\be}{\begin{equation}}
\newcommand{\ee}{\end{equation}}
\newcommand{\bea}{\begin{eqnarray}}
\newcommand{\eea}{\end{eqnarray}}
\newcommand{\ba}{\begin{array}}
\newcommand{\ea}{\end{array}}

\begin{document}

\begin{titlepage}
\begin{flushright}
July, 2008\\
\end{flushright}

\vfill

\begin{center}
{\large \bf
Quantum gravity equation in large $N$ Yang-Mills quantum mechanics}
\vfill
{
Toshihiro Matsuo$^{*,}$\footnote{\tt tmatsuo@het.ph.tsukuba.ac.jp},
}
{
Dan Tomino$^{\star,}$\footnote{\tt dan@home.phy.ntnu.edu.tw},
}
{
Wen-Yu Wen$^{\dagger,\spadesuit}$\footnote{\tt steve.wen@gmail.com}
}
and
{
Syoji Zeze$^{\dagger,\heartsuit}$\footnote{\tt s-zeze@seiryou-h.akita-c.ed.jp}
}
\vfill
$^{*}$ {\it
Graduate School of Pure and Applied Sciences, University of Tsukuba, \\
Ibaraki 305-8571, Japan} \vskip 0.4truecm 
$^{\star}$ {\it
Department of Physics,
National Taiwan Normal University,\\
Taipei City, 116, Taiwan } \vskip 0.4truecm 
$^{\dagger}$ {\it
Department of Physics and Center for Theoretical Sciences,
National Taiwan University,\\
Taipei, 10617, Taiwan} \vskip 0.4truecm
$^{\spadesuit}$ {\it
Leung Center for Cosmology and Particle Astrophysics,\\
National Taiwan University, \\
Taipei, 10617, Taiwan} \vskip 0.4truecm 
$^{\heartsuit}$ {\it
Yokote Seiryo Gakuin High School,\\
Akita 013-0041, Japan} \vskip 0.4truecm 

\end{center}

\vfill
\begin{abstract}
We propose a new interpretation of the BFSS large $N$ matrix quantum mechanics analogous to a novel interpretation of the IKKT matrix model where infinitely large $N$ matrices act as  differential operators in a curved space.  In this picture, the Schr\"odinger equation in the BFSS model is regarded as the Wheeler-DeWitt equation which determines the wave function of universe.  An explicit solution of wave function is studied in a simple two-dimensional minisuperspace model.  
\end{abstract}
\vfill

\end{titlepage}

\setcounter{footnote}{0}

\section{Introduction}
\label{intro}
\setcounter{equation}{0}
In matrix model proposals for non-perturbative definition of M-theory (the BFSS model \cite{Banks:1996vh}) and superstring theory (the IKKT model \cite{Ishibashi:1996xs}), realization of quantum theory of gravity has always been a major concern.
String theory in its perturbative formulation contains graviton excitation and it gives a consistent quantum theory of gravity. 
M-theory is defined as the strong coupling limit of type IIA string theory and should include the eleven dimensional supergravity as low energy effective theory.  It has been expected that their quantum aspects can be captured in terms of the IKKT and the BFSS matrix models.

However, it is not easy to identify these matrix models with quantum gravity, because general covariance is not manifest in their formulations.  Rather, they are formulated as large $N$ limit of gauge theories; namely dimensional reductions of ${\cal N}=1$ super Yang-Mills theory in ten dimensions down to 0+0 dimension (IKKT) or 0+1 dimension (BFSS).  
Traditionally, these theories have been interpreted as a systems of infinitely many D-branes.  In such interpretation, the gravitational effect appears rather indirectly through the loop effect of open strings stretched between each D-branes. Such idea has already been examined, and it was found that the loop effect reproduces the graviton exchange between two D-objects \cite{Banks:1996vh, Ishibashi:1996xs}.  The graviton exchange was extensively investigated further in higher orders of loop expansion and derivative terms, and correspondence between the matrix theory and supergravity was reported \cite{Taylor:2001vb, Okawa:1998pz}.  As another attempt the induced action of gravity by a loop effect in the matrix model is discussed in \cite{Grosse:2008xr}.  It has been a difficult task, however, 
to find a way of realizing quantum gravity in terms of matrix model without relying on any perturbation\footnote{
As for recent studies on string theory aspects of the models, a light-cone superstring action has been derived from IIB matrix model \cite{Kitazawa:2007gp}, and vertex operators and scattering amplitudes of superstring are reproduced in \cite{Kitazawa:2007um}. }.

In recent years, an alternative identification of gravitational degrees of freedom in matrix model have been proposed by Hanada, Kawai and Kimura \cite{Hanada:2005vr}.  These authors showed that a matrix can be identified with a covariant derivative in a curved spacetime in large $N$ limit.  Once we apply this identification to the IKKT model, for instance, we find that the matrix equation of motion gives Einstein equation in vacuum.  In this paper we will call such an identification in matrix model as ``HKK interpretation'' for convenience.  Following this approach, we do not need any loop calculation, and the Einstein equation is derived quite straightforwardly.  Such feature is quite attractive even though there is still subtlety of regularization due to infinite size of matrices.  

Past studies along this line have been devoted to the IKKT matrix model \cite{Hanada:2006gg, Hanada:2006ei,Furuta:2006kk}.  In this paper we address an  application of the HKK interpretation to the BFSS matrix model.\footnote{Although we study the large $N$ (bosonic) Yang-Mills quantum mechanics with $d$ matrices, still we call it the ``BFSS" model. Similarly we call the large $N$ Yang-Mills reduced model with $d+1$ matrices the ``IKKT" model.}
This academic application immediately gives us some interesting results.  We recall that the BFSS matrix model is a quantum mechanics whose dynamical variable are time-dependent matrices.  It is, therefore, natural to expect that the Schr\"odinger equation describes quantum evolution of spacetime.  This is remarkable advantage of our formalism in contrast to the IKKT matrix
model whose matrices are time independent.
The Schr\"odinger equation becomes in the HKK prescription:
\begin{equation}
i\frac{\partial}{\partial t}\Psi(t, X)=H(X,\partial/\partial X)\Psi(t,X)\quad
\longrightarrow \quad i\frac{\partial}{\partial t}\Psi(t,\nabla)=H(\nabla, \partial/\partial \nabla)\Psi(t, \nabla),\nonumber 
\end{equation}
where $\nabla$ is a covariant derivative in a curved space described by the vielbein and the spin connection.  A wave function obtained by solving the above equation describes quantum evolution of the spacetime.  
We will also address the meaning of expressions such as $\Psi=\Psi(t,\nabla)$ and $\partial/\partial \nabla$ which look somewhat symbolic.

This paper is organized as follows.  
In section \ref{HKK-BFSS}, after reviewing the HKK interpretation and the quantization procedure of the BFSS model applicable to finite $N$ case, we apply the interpretation to the BFSS model and obtain a Hamiltonian in terms of the HKK interpretation. 
The Schr\"odinger equation is derived straightforwardly from the Hamiltonian after quantization.  
This process is in contrast to traditional approaches of quantum gravity where the Schr\"odinger equation (the Wheeler-DeWitt equation) is obtained as a Hamiltonian constraint.  
In section \ref{MSSM}, we examine a minisuperspace model in two dimensions as an explicit example.  
We find that the time independent Schr\"odinger equation can be solved analytically. 
We also obtain a time dependent wave function by constructing a wave packet.  Expectation values and dispersions of various geometrical quantities such as metric and torsion are evaluated. Finally we discuss physical implication of our result.

\section{HKK interpretation of BFSS model}
\label{HKK-BFSS}
\setcounter{equation}{0}
\subsection{HKK interpretation for large $N$ matrices}
\label{HKK}
We begin with a brief review on the HKK interpretation of matrix models proposed by \cite{Hanada:2005vr} where the authors argued how to interpret large $N$ matrices as covariant derivatives in curved spacetime.  Their formalism can be applied to arbitrary mainfold at least formally, and a case of some compact manifolds had been studied in detail\cite{Hanada:2006ei}.  In their proposal, a large $N$ matrix is identified with an operator which acts on a section of vector bundle $E_{reg}$ over $D(=1+d)$ dimensional curved manifold $M$ endowed with structure group $G$, where $G=Spin(1,d)$ is local Lorentz group\footnote{ 
The section of a fiber bundle $\Gamma(E_{reg})$ is a set of smooth maps from a coordinate patch of $M$ to some vector space $V_{reg}$.  $V_{reg}=\{f\;:\;G\rightarrow {\cal C}\}$ is reducible and has following decomposition
\begin{eqnarray*}
V_{reg}=\oplus_r\underbrace{(V_r\oplus\cdots\oplus V_r)}_{d_r},
\end{eqnarray*}
where $V_r$ is the space of irreducible representation $r$ of $G$,  and $d_r$ is its dimension.
}. 
The relation between large $N$ matrix $X_{(a)}$ and the covariant derivative $\nabla_{a}$ on a manifold $M$ is given as 
\begin{eqnarray}
X_{(a)} =i R_{(a)}^{\quad a}(g^{-1}) \nabla_{a}.
\label{exp1}
\end{eqnarray}
The matrix $R_{(a)}^{\quad a}(g^{-1})$ is the vector representation of local Lorentz group $G$ $(g \in G)$.
While the index $a$ transforms as Lorentz vector with respect to the
action of $G$ on $\Gamma(E_{reg})$, the other index $(a)$ remains unchanged which is a index of global $SO(D)$.  This feature enables us to identify ($\ref{exp1}$) as an endomorphism on $\Gamma(E_{reg})$, i.e.,  matrix acts on this space.  In this formalism, the covariant derivative $\nabla_{a}$ is interpreted as a linear map from $\Gamma(E_{reg})$ to $TM\otimes\Gamma(E_{reg})$, where $TM$ is the tangent bundle on $M$.  It can be written explicitly as
\begin{equation}
\nabla_{a}={e_{a}}^{\mu}(\partial_{\mu}+{\omega_{\mu}}^{bc}{\cal O}_{bc}) ,
\end{equation}
where $e$ and $\omega$ are vielbein and spin-connection respectively, and $\cal{O}$ is Lorentz generator whose explicit form depends on a representation on which it acts.

\subsection{BFSS - matrix quantum mechanics}
\label{BFSS}
Before applying the HKK prescription to the BFSS model, let us recall quantization of the model in a way that is convenient for later argument.
The bosonic part of the BFSS action in Lorentzian signature is given by
\begin{eqnarray}
S_{BFSS}
= \int dt L
=\int dt \, tr \left(
-\frac{1}{2}  [D_{0},X_{(i)}][D^{0},X^{(i)}]
+\frac{1}{4} [X_{(i)}, X_{(j)}] [X^{(i)}, X^{(j)}]
\right),\label{BFSSaction}
\end{eqnarray}
where $D_{0}=\partial_{t} + i A_{0}$ is the gauge covariant derivative.  The indices $(i), (j)$ run only in $d$ spatial directions.
We shall work in the $A_0=0$ gauge such that $[D_0, X]=\partial_t X$.
We employ the matrix notation as
\bea
X^{(i)}_{ab}(t)=x^{(i)}_A(t) t^A_{ab} ,
\label{Xexpansion}
\eea
where $t_{ab}^A$ is the generator of $SU(N)$ algebra in adjoint representation so that $a$ and $b$ run from $1$ to $N$ and indices $A$ from $1$ to $N^2-1$.  $x^{(i)}_{A}$ is the degrees of freedom and $t_{ab}^A$ are the basis.
The conjugate momentum of $X_{(i)}$ is defined as
\begin{eqnarray}
\Pi^{(i)}_{ab}\equiv \frac{\partial L}{\partial \dot{X}_{(i)ab}}=\dot{X}^{(i)}_{ab} ,
\label{sc2}
\end{eqnarray}
where $\dot{X}$ denotes $\partial_t X$.
The Legendre transformation gives us the BFSS Hamiltonian
\begin{eqnarray}
H&=& tr\left\{\frac{1}{2}\Pi_{(i)}\Pi^{(i)}-\frac{1}{4} [X_{(i)},X_{(j)}][X^{(i)},X^{(j)}] \right\} .
\label{sc3.0}
\end{eqnarray}
The gauge constraint arising from our choice of $A_{0}=0$ gauge is
\begin{eqnarray}
[X^{(i)}, \Pi_{(i)}]=0.
\label{sc4}
\end{eqnarray}
The Poisson bracket of $X_{(i)}$ and $\Pi^{(j)}$ is written as
\begin{eqnarray}
\left\{ {X_{(i)}}_{ab}, \Pi^{(j)}_{cd} \right\}_{PB} ={\delta_{(i)}}^{(j)}h_{AB}t^A_{ab}t^B_{cd} ,
\label{poisson3}
\end{eqnarray}
where $h_{AB}$ is the inverse of the metric $h^{AB}=t^A_{ab}t^B_{ba}$.
Upon quantization the Poisson bracket becomes the canonical commutator, and $\Pi^{(i)}$ acts as $-i\partial/\partial X_{(i)}$ on the wave function $\Psi(X)$.
The Schr\"odinger equation then becomes
\begin{eqnarray}
i\frac{\partial}{\partial t}\Psi(X)&=&
tr\left( -\frac{1}{2} \frac{\partial}{\partial X_{(i)}}\frac{\partial}{\partial X^{(i)}} -\frac{1}{4} [X_{(i)},X_{(j)}] [X^{(i)},X^{(j)}]  \right)\Psi(X) .
\label{sc5}
\end{eqnarray}
Having reviewed the matrix quantum mechanics, we will apply the HKK prescription to the above argument in the next subsection.

\subsection{Applying HKK to BFSS Shr\"odinger equation} 
\label{HKK_to_BFSS}
We set the matrix $X_{(i)} (t)$ as a covariant derivative:  
\begin{eqnarray}
X_{(i)}= i{R_{(i)}}^i\nabla_i= i{R_{(i)}}^i({{e_i}}^{I}\partial_{I}+{{\omega_i}}^{jk}{\cal O}_{jk}), 
\qquad {\omega_i}^{jk}={e_i}^I {\omega_I}^{jk},
\label{kin1}
\end{eqnarray}
where ${R_{(i)}}^{j} (\hat{g}^{-1})$ belongs to $\hat{G}=Spin(d)$ vector representation and all the indices run only in $d$ spatial dimensions.  
Note that the expression (\ref{kin1}) is analogous to (\ref{Xexpansion}) if we regard $\partial_I$ and ${\cal O}_{jk}$ as basis, and the vielbein and the spin connection as degrees of freedom.  
We shall make more comments on this formal argument in Appendix \ref{momentum}.
We also require each index in (\ref{kin1}) to transform appropriately in $d$ dimensions.
Thus the covariant derivative in (\ref{kin1}) is associated with a $d$ dimensional manifold for each $t$.  One may wonder whether this manifold corresponds to a time slice of some $d+1$ dimensional manifold whose timelike Killing vector is given by $\partial_{t}$.  
Indeed, in Appendix \ref{BFSSintoIKKT}, we demonstrate that one can choose a local frame equipped with metric $-dt^{2} +h_d(t,x)_{IJ}dx^{I}dx^{J}$ such that $X_{(i)} (t)$ and $D_{0}$ are embedded into the $d+1$ dimensional IKKT model, although we do not pursue along this line of reasoning here.  

The commutator becomes\!
\footnote{The representation matrix ${R_{(i)}}^{j} (\hat{g}^{-1})$ can be moved into left side of the covariant derivative $\nabla_{i}$. It was explained in \cite{Hanada:2005vr} in detail.}   
\begin{eqnarray}
[X_{(i)},X_{(j)}] =-{R_{(i)}}^i{R_{(j)}}^j[\nabla_i,\nabla_j],
\label{po2}
\end{eqnarray}
with
\begin{eqnarray}
[\nabla_i, \nabla_j] &=& [e_{i}^{\;I}\nabla_{I}, e_{j}^{\;J}\nabla_{J}]
= {T_{ij}}^{K}\partial_{K}+R_{ij}^{\;\; kl}{\cal O}_{kl},
\end{eqnarray}
where
\begin{eqnarray}
{T_{ij}}^{K}&\equiv& e_{[i}^{\;L}\partial_{L} e_{j]}^{\;K} + \omega_{[i\;j]}^{\quad K}, \label{pot3}\\
{R_{ij}}^{kl}&\equiv&
{{\cal R}_{ij}}^{kl} + {T_{ij}}^K {\omega_K}^{kl} 
\nonumber \\
&=&
 e_{[i}^{\;K}\partial_{K}{\omega_{j]}}^{kl}
+\omega_{[i}^{\;\;km}{\omega_{j]}}^{ml}
+ \omega_{[i\;j]}^{\quad m}{\omega_{m}}^{kl}.
\label{pot4}
\end{eqnarray}
Here ${T_{ij}}^K$ is the torsion and ${{\cal R}_{ij}}^{kl}$ is the Riemann curvature tensor.  
The reader may refer to Appendix \ref{Torsion_Curvature} for derivation. 
If the torsion free condition ${T_{ij}}^K=0$ is satisfied, then ${e_i}^I$ and ${\omega_i}^{jk}$ are no longer independent of each other.  We have learnt that the first proposal discussed in \cite{Hanada:2005vr} considered torsion free case for simplicity. 
In general, however, matrix configurations do not necessary to satisfy the torsion free condition. 
We remark that torsion in the IKKT model has been discussed in \cite{Furuta:2006kk}.

Next we apply the HKK prescription to the conjugate momentum (\ref{sc2}). We find
\begin{eqnarray}
\Pi^{(i)}= \dot{X}^{(i)}=i\delta^{(i)(l)}{R_{(l)}}^i
({\dot{e}_{i}}^{\;\;I}\partial_{I}+{\dot{\omega}_i }^{\;\;jk}{\cal O}_{jk}).
\label{kin2}
\end{eqnarray}
In (\ref{kin1}) and (\ref{kin2}), only ${e_{i}}^{I}$ and ${\omega_i}^{jk}$ and their time derivatives are dynamical, while $\partial_I$ and ${\cal O}_{jk}$ are independent of the choice of geometry.  Therefore it is convenient to  introduce a phase space defined
 by $\{{e_{i}}^{I}$, ${\omega_i}^{jk}\}$ and their conjugate momenta
\begin{eqnarray}
\{ {\pi^{e}}_{i}^{\;\;I} =\dot{e}_{i}^{\;\;I}, \quad 
 { \pi^{\omega}}_{i}^{\;jk} =\dot{\omega}_{i}^{\;\;jk} \},
\label{kin3}
\end{eqnarray} 
where the Poisson brackets are defined by
\begin{equation}
\{e_{i}^{\;I} (x), {\pi^e}_{j}^{\;J} (y) \}_{PB}  =
 \delta_{ij}\delta^{IJ} \delta^{d} (x-y), \quad
 \{\omega_{i}^{\; j k} (x), {\pi^{\omega}}_{l}^{ \; m n}(y)   \}_{PB} =
 \delta_{il}\delta^{jm} \delta^{kn}\delta^{d} (x-y).
\label{poisson1}
\end{equation}
Then using the above definition, the Poisson bracket between
$X_{(i)}$ and $\Pi_{(j)}$ can be evaluated as
\begin{equation}
\left\{ X_{(i)}(x)_{\alpha\beta}, \Pi^{(j)}(y)_{\gamma\delta} \right\}_{PB} =
{\delta_{(i)}}^{(j)}\delta^{d}(x-y)\left(
-\partial_{I}\partial^{I}{\bf 1}_{\alpha\beta}{\bf 1}_{\gamma\delta}  
-({\cal O}_{lm})_{\alpha\beta}({\cal O}^{lm})_{\gamma\delta})
\right), \label{poisson2}
\end{equation}
where we write matrix indices along fiber direction explicitly in Greek subscripts. 
We have also used the orthogonal property ${R_{(i)}}^{k}{R_{(j)}}^{k}=\delta_{(i)(j)}$.
Note that there is a ``factor'' in (\ref{poisson2}) which multiplies 
$\delta_{(i)}^{(j)}\delta^{d}(x-y)$. We also learnt that 
similar factor also appears in (\ref{poisson3}).
In fact, such analogy between (\ref{poisson2}) and (\ref{poisson3}) can be seen by identifying $\partial_{I}$ and $\mathcal{O}_{ij}$ as $t^{A}_{ab}$ in (\ref{Xexpansion}), but we can not put forward this analogy further since $t^{A}$ is a matrix but $\partial_{I}$ and $\mathcal{O}_{ij}$ are not. Parts of these combining with ${R_{(i)}}^i$ may provide corresponding basis.
See Appendix \ref{momentum}.

Having defined the phase space, we are ready to construct Hamiltonian which 
governs dynamics of our model. 
It can be obtained by applying (\ref{kin1}) and (\ref{kin2}) to the 
BFSS Hamiltonian defined by (\ref{sc3.0}).  The trace in (\ref{sc3.0})
should be performed over a complete set of function in $C^{\infty} (E_{prin})$. 
In this way we have
\begin{eqnarray}
H&=&-\int d^dx \sqrt{h_d}\;tr_{\hat{g}}
\left\{ \frac{1}{2}(\pi_{i}^{e\;K}\partial_{K})(\pi_{i}^{e\;L}\partial_{L})
+\frac{1}{2}(\pi_{i}^{\omega\;jk}{\cal O}_{jk})(\pi_{i}^{\omega\;lm}{\cal O}_{lm}) 
\right. 
\nonumber \\
&&\hspace{30mm}
\left.
+\frac{1}{4}({T_{ij}}^{K}\partial_{K})({T_{ij}}^{L}\partial_{L})
+\frac{1}{4}({R_{ij}}^{ij}{\cal O}_{ij})({R_{ij}}^{kl}{\cal O}_{kl}) \right\},
\label{hal1}
\end{eqnarray}
where we have used (\ref{nhkk4}) for $tr$ in the large $N$ limit.
The operation $tr_{\hat{g}}$ is defined as $tr_{\hat{g}}F=\int d\hat{g}\langle x,\hat{g}|F|x,\hat{g}\rangle$ with a Haar measure $d\hat{g}$ for $\hat{g}$.
The matrix ${R_{(i)}}^i$ has disappeared thanks to its orthogonality. 
There is no cross term because $tr_{\hat{g}}{\cal O}=0$.
There are the trace operations such as $\int d^dx \partial^2$ and $tr_{\hat{g}} {\cal O}^2$ and they formally diverge due to tracing over a regular representation with infinite dimensions. 
Here we simply assume that we have employed a suitable regularization procedure, which could be the heat kernel regularization or elsewhere suggested in \cite{Hanada:2006ei}. 

Having obtained classical Hamiltonian, we can now quantize the system by replacing the Poisson brackets in (\ref{poisson1})
with commutators.  We shall work in a representation such that
$e$ and $\omega$ becomes diagonal.  With this choice,
 $\pi_{i}^{e\;I}$ and $\pi_{i}^{\omega\;jk}$ are promoted into operators,
\begin{eqnarray}
\pi_{i}^{e\;I}=-i\frac{\delta}{\delta e_{i}^{\;I}}, \quad
\pi_{i}^{\omega\;jk}=-i\frac{\delta}{\delta \omega_{i}^{\;\;jk}}.
\label{sc6}
\end{eqnarray} 
Note that there is a problem of operator ordering due to the presence of the determinant factor in the Hamiltonian.
Choosing an ordering prescription such that the Hamiltonian preserves hermiticity, the Schr\"odinger equation (\ref{sc5}) becomes 
\begin{eqnarray}
i\frac{\partial }{\partial t}\Psi=\int d^dx \sqrt{h_d}\;tr_{\hat{g}}
\Bigg\{ \frac{1}{\sqrt{h_d}}  \frac{1}{2}
(\frac{\delta}{\delta e_{i}^{K}}\partial_{K})\sqrt{h_d}(\frac{\delta}{\delta e_{i}^{L}}\partial_{L})
+\frac{1}{2}
(\frac{\delta}{\delta {\omega_{i}}^{jk}}{\cal O}_{jk}) (\frac{\delta}{\delta {\omega_{i}}^{lm}}{\cal O}_{lm})
\nonumber \\
-\frac{1}{\sqrt{h_d}}\frac{1}{4}({T_{ij}}^{K}\partial_{K})\sqrt{h_d}({T_{ij}}^{L}\partial_{L})
-\frac{1}{4}({R_{ij}}^{kl}{\cal O}_{kl})({R_{ij}}^{mn}{\cal O}_{mn}) \Bigg\} \Psi .
\label{sc8}
\end{eqnarray}
In addition, the gauge constraint (\ref{sc4}), while imposing on the wave function, takes the following form
\begin{eqnarray}
(\tilde{T}^{K}\partial_K+\tilde{R}^{ij}{\cal O}_{ij})\Psi=0,
\label{sc9}
\end{eqnarray}
where
\begin{eqnarray}
\tilde{T}^{K}&\equiv& e_i^{\;L}(\partial_{L} \dot{e}_{i}^{\;K}) - \dot{e}_i^{\;L}(\partial_{L} e_{i}^{\;K})
+\omega_{ii}^{\;\; l}\dot{e}_{l}^{\;K}-\dot{\omega}_{ii}^{\;\; l}e_{l}^{\;K}, \label{sc10}
\nonumber\\
\tilde{R}^{ij}&\equiv&
 e_{k}^{\;K}\partial_{K}\dot{\omega}_{k}^{\;ij}- \dot{e}_{k}^{\;K}\partial_{K}\omega_{k}^{\;ij}
+\omega_{l}^{\;ik}\dot{\omega}_{l}^{\;kj}
-\dot{\omega}_{l}^{\;ik}\omega_{l}^{\;kj}
+ \omega_{ll}^{\;\;k}\dot{\omega}_{k}^{\;\;ij}-\dot{\omega}_{ll}^{\;\; k}\omega_{k}^{\;\;ij}. 
\label{sc11}
\end{eqnarray}
The equation (\ref{sc8}) together with the constraint (\ref{sc9}) dictate the evolution of quantum system with degrees of freedom given by vielbein $e$ and spin connection $\omega$, which in turn determine gravity in the classical level. 
To summarize, we regard that the Schr\"odinger equation (\ref{sc8}) and the gauge constraint (\ref{sc9}) are the equations of quantum gravity which is realized in the BFSS quantum mechanics at large $N$ limit.

In the case the Hamiltonian in (\ref{sc8}) is time independent, by substituting $\Psi(t,x)=e^{-iEt}\Psi_E(x)$ we obtain the time independent Schr\"odinger equation
\begin{eqnarray}
&&
\int d^dx \sqrt{h_d}\;tr_{\hat{g}}
\Bigg\{ \frac{1}{\sqrt{h_d}}  \frac{1}{2}
(\frac{\delta}{\delta {e_{i}}^{K}}\partial_{K})\sqrt{h_d}(\frac{\delta}{\delta {e_{i}}^{L}}\partial_{L})
+\frac{1}{2}
(\frac{\delta}{\delta {\omega_{i}}^{jk}}{\cal O}_{jk}) (\frac{\delta}{\delta {\omega_{i}}^{lm}}{\cal O}_{lm})
\nonumber \\
&&\hspace{5mm}
-\frac{1}{\sqrt{h_d}}\frac{1}{4}({T_{ij}}^{K}\partial_{K})\sqrt{h_d}({T_{ij}}^{L}\partial_{L})
-\frac{1}{4}({R_{ij}}^{kl}{\cal O}_{kl})({R_{ij}}^{mn}{\cal O}_{mn})\Bigg\} \Psi_E=E \Psi_E.
\label{sc12}
\end{eqnarray}
The appearance of this equation, at first sight, is analogous to the Wheeler-DeWitt equation in canonical quantum gravity \cite{DeWitt:1967yk}.  One might, however, question their similarity mainly because the Wheeler-DeWitt equation is seen as a  Hamiltonian constraint without time evolution due to reparametrization invariance. 
Nevertheless it is known that the Schr\"odinger equation can be rewritten as a Hamiltonian constraint through time reparametrization, i.e. $t=t(\tau)$ \cite{Brown:1989ne}.  In this sense the equation (\ref{sc8}) and (\ref{sc12}) indeed play the same role as the Wheeler-DeWitt equation does. 
Here we briefly recall the argument.
We introduce the lapse function $N(\tau)$ such that 
\bea
dt(\tau)=N(\tau)d\tau
\eea
in the action  (\ref{BFSSaction}). 
The variable $t(\tau)$ is regarded as an independent variable with equal footing as $X(\tau)$.
Introducing a Lagrange multiplier $\pi$ which plays the role of the conjugate momentum with respect to $t(\tau)$, we have
\bea
S= Tr \int d\tau \left[
{1 \over 2N(\tau)}\left({dX^{(i)}(\tau) \over d\tau}\right)^2-{N(\tau)\over 4}[X^{(i)},X^{(j)}]^2+\pi\left({d t(\tau) \over d \tau}-N(\tau)\right)
\right].
\eea
The conjugate momentum with respect to $X^{(i)}(\tau)$ becomes
\bea
P^{(i)}={1 \over N}{dX^{(i)}\over d\tau},
\eea
then we obtain
\bea
S=Tr \int d\tau\left[
P_{(i)} \dot{X}^{(i)} +\pi \dot{t}-N C_H
\right],
\eea
where
\bea
C_H={1\over 2}P_{(i)}^2+{1\over 4}[X^{(i)},X^{(j)}]^2+\pi.
\eea
In this way we obtain the canonical system with variables $(X,t)$ whose conjugate momenta are $(P,\pi)$ respectively and the Hamiltonian is given by $NC_H$.
We obtain a constraint $C_H=0$ by the variation of $N$.
Clearly the action is invariant under the reparametrization $\tau \to \tau'=\tau'(\tau)$ in which $N \to N'(\tau')=d\tau/d\tau' N(\tau)$, and all the others transform as scalars.
The Lagrange multiplier $\pi$ plays a role of $E$ in (\ref{sc12}).

Now one can consider the WKB approximation as a semi-classical limit of the large $N$ BFSS model. In the WKB approximation, a wave function is governed by a classical action evaluated at saddle points, which satisfy the classical equations of motion. In the present case we have
\begin{eqnarray}
[D^0,[D_0,X_{(i)}]]+[X^{(j)},[X_{(j)},X_{(i)}]]&=&0,
\label{cqe1}\\
{}[X^{(i)},[X_{(i)},D_{0}]]&=&0.
\label{cqe2}
\end{eqnarray}
By using the HKK interpretation, they become (see Appendix \ref{EOMappendix} for derivation) 
\begin{eqnarray}
\ddot{e}_{i}^{\;\;I}
+(\nabla^k{T_{ki}}^{j})e_{j}^{\;\;I}+\eta^{lk}{T_{ki}}^{m} {T_{lm}}^{I}
-{T_{ji}}^{k}\omega^{k}_{\;jI}+{R_{i}}^{I} &=&0, \\
 \ddot{\omega}_{ikl}
-\ddot{e}_{i}^{\;\;I}e_{I}^{\;\;j}\omega_{jkl}
+{T_{mi}}^{n}(R^{m}_{\;nkl}-\eta^{mp}{T_{pm}}^{j}\omega_{jkl})+\nabla^{j}(R_{jikl}-{T_{ij}}^{m}\omega_{mkl})&=&0.
\end{eqnarray}
Hence the semi-classical limit of the large $N$ BFSS model describes a theory of gravity with/without torsion, even though the Hamiltonian itself looks very different from that in the Einstein theory.  In other words, quantum nature of the BFSS model could be
different from what has been derived from quantization of the ordinary Einstein-Hilbert action.
However, we should remind that above consideration is somewhat naive therefore has to be refined at least with respect to the following points: 
the first point is that although the form of equation keeps the form of Einstein equation, general covariance is lost in our application of HKK to the BFSS model, namely complete classical Einstein equation appears only in spatial directions. 
The second point is that in the quantum mechanics the time evolution of expectation value of an operator ${\cal A}$ obeys Hamilton's equation,  
\begin{eqnarray}
\frac{d}{dt}\langle{\cal A}\rangle=i\langle[H,{\cal A}] \rangle.
\label{hamileq}
\end{eqnarray}
Then expectation values of the position operator $\hat{x}$ and the momentum operator $\hat{p}$ in quantum mechanics obey classical equation of motion.
Here we have additional metric determinant in the large $N$ Hamiltonian and its time evolution makes time dependence of expectation values depart from classical matrix equations of motion. 
These discrepancies from classical Einstein theory will be significant where spacetime is far from the flat one without torsion.  
%

\section{Minisuperspace model  }
\label{MSSM}
\setcounter{equation}{0}
In this section, we will examine our new interpretation in a minisuperspace model as an explicit example.  To be precise,  we will consider an exactly calculable toy model of two dimensional universe to clarify our proposal given in the previous section.  We start with a simple (gauged) one matrix model without potential term:
\begin{eqnarray}
{\cal L}=\frac{1}{2}tr[D_0, X]^2.
\label{mlag}
\end{eqnarray}
It seems to be almost free, but the metric determinant factor which appears after taking large $N$ limit gives nontrivial dynamics. 
Assuming our toy model of universe is given by the following metric:
\begin{eqnarray}
ds^2=-dt^2+\frac{1}{y(t)^2}dx^2.
\label{mhkk2}
\end{eqnarray}
This is a two dimensional version of the Robertson-Walker universe. 
In the assumption here the function $y(t)$ only depends on time. 
It makes our setting to be a tractable problem. 
Following the HKK interpretation, we obtain the covariant derivative:
\begin{eqnarray}
D_0=i\frac{\partial}{\partial t},  \quad X=iy(t) \frac{\partial}{\partial x}, 
\label{mhkk1}
\end{eqnarray}
as well as the time independent Schr\"odinger equation:
\begin{eqnarray}
\left[-A\frac{\partial}{\partial y}\frac{1}{y}\frac{\partial }{\partial y}\right]\Psi=E\Psi,&& \nonumber\\
A=-tr_{\hat{g}}\int dx (\partial_x)^2 .
\label{msc1}
\end{eqnarray}
Here $A$ is a positive and divergent quantity which is expected to be regularized by some regularization procedure like the heat kernel regularization proposed in \cite{Hanada:2006ei}, and in the following we will treat $A$ as if it is a finite quantity.
The factor $1/y$ sitting between the derivatives is originated from the metric determinant factor in the Hamiltonian.
The gauge constraint (\ref{sc4}) on the other hand is trivially satisfied in this setting. 
From (\ref{msc1}), we obtain
\begin{eqnarray}
\left[\frac{d}{dy} \frac{1}{y} \frac{d}{dy} +\epsilon\right] \Psi=0, \quad 
\epsilon=E/A.
\label{msc3}
\end{eqnarray}
A solution to (\ref{msc3}) for $\epsilon>0$ \footnote{It corresponds to plane wave type solution in the BFSS quantum mechanics.} is found to be  
\begin{eqnarray}
\Psi_{\epsilon}=\frac{1}{\sqrt{3}}yJ_{\pm\frac{2}{3}}\left(\frac{2}{3}\sqrt{\epsilon}y^{\frac{3}{2}}\right),
\label{sol1}
\end{eqnarray}
where $J_{\nu}(z)$ is the Bessel function.  The wave function $\Psi_\epsilon$ is orthogonal 
\begin{eqnarray}
\int_{0}^{\infty} \Psi_{\epsilon}(y)\Psi_{\epsilon'}(y)dy=\delta(\epsilon-\epsilon').
\label{sol2} 
\end{eqnarray}
We have provided a proof for the orthogonality in appendix \ref{orthogonality}.

Using this orthogonal basis (\ref{sol1}), time dependent wave packet is constructed as follows:
\begin{eqnarray}
\Psi(t,y)=\int^{\infty}_{0} d\epsilon \;C(\epsilon)e^{-iA\epsilon t}\Psi_{\epsilon}(y) ,
\label{wp1} 
\end{eqnarray}
where $C(\epsilon)$ is arbitrary function that is normalizable. Here we take the following Boltzmann form as an example
\begin{eqnarray}
C(\epsilon)=\sqrt{\frac{(2\beta)^{\frac{5}{3}}}{\Gamma(5/3)}}\epsilon^{\frac{1}{3}}e^{-\beta \epsilon},
\label{wp2}
\end{eqnarray}
such that normalization condition $\int d\epsilon C(\epsilon)^2=1$ is satisfied.  $\beta$ is a parameter to characterize the wave packet.  Then equation (\ref{wp1}) becomes 
\begin{eqnarray}
\Psi(t,y)=\sqrt{\frac{(2\beta)^{\frac{5}{3}}}{\Gamma(5/3)}} 
\frac{ y^2 } {3^{\frac{7}{6}}(\beta+iAt)^{\frac{5}{3}}}
\exp\left[-\frac{y^3}{9(\beta+iAt)}\right] .
\label{sol9}
\end{eqnarray}
The wave packet becomes zero as $t\rightarrow  \pm\infty$ or $y\rightarrow \infty$ and its peak is located at $(t,y)=(0,(15\beta)^{1/3})$. 
We observe that the wave packet exponentially decays for large $y$ although the metric (\ref{mhkk2}) shows a singularity at $y\to \infty$.  However there is no pathology about this because the wave function itself is regular and smooth for all values of $y$.

The time evolution can be seen by evaluating the following expectation value:
\begin{eqnarray}
\langle\;y^m\;\rangle=\int_{0}^{\infty} dy\; \Psi^{*}(t,y)y^m\Psi(t,y)=\frac{\Gamma(\frac{5}{3}+\frac{m}{3})}{\Gamma(\frac{5}{3})} 
\left({9\over 2}\right)^{\frac{m}{3}}
\left[\frac{\beta^2+A^2t^2}{\beta}\right]^{\frac{m}{3}} ,
\label{exv1}
\end{eqnarray}
where we have used (\ref{sol9}). 
One can easily calculate various expectation value of operators based on (\ref{exv1}).  
To our most concern, the scale factor is found to be 
\begin{eqnarray}
\langle\;g_{xx}\;\rangle = \langle\;y^{-2}\;\rangle=
\frac{1}{\Gamma(\frac{5}{3})}
\left({2\over 9}\right)^{\frac{2}{3}}
\left[\frac{\beta}{\beta^2+A^2t^2}\right]^{\frac{2}{3}}.
\label{exv3}
\end{eqnarray}
We observe that around $t=0$ the expectation value of size of universe keeps a constant value, and late time behavior is $\sim t^{-2/3}$. The universe shrinks to zero size as $t\rightarrow \infty$.
We can also calculate expectation value of torsion. In this model the non-vanishing torsion operator is given by
\begin{eqnarray}
T=-i\frac{d}{dy} ,
\label{exv4}
\end{eqnarray}
and one obtains 
\begin{eqnarray}
\langle\;T\;\rangle= 
\frac{\Gamma(\frac{4}{3})}{\Gamma(\frac{5}{3})}
\left({4\over 3}\right)^{\frac{2}{3}}
\frac{\beta^{-\frac{3}{2}}At}{(\beta^2+A^2t^2)^{\frac{1}{3}}} ,
\label{exv6}
\end{eqnarray}
which shows that $\langle\;T\;\rangle\sim t$ for small $|t|$ and $|\langle\;T\;\rangle| \sim |t|^{1/3}$ for large $|t|$.  
Since the torsion gives kinetic energy, the energy density stored in this universe is not zero. In fact, it can be calculated as
\begin{eqnarray}
\langle\; \epsilon \;\rangle = \frac{5}{6\beta}.  
\label{exv8}
\end{eqnarray}

We have mentioned that the relation to the conventional torsion free gravity is obtained by imposing torsion free condition to the wave function, say $T\Psi=0$. However, it is easy to see that it is only satisfied by trivial wave functions thus the torsion free condition gives trivial solution in this model.

It is also interesting to see dispersion of an operator ${\cal A}$, given by
\begin{eqnarray*}
\Delta^2_{{\cal A}}&\equiv& \langle ({\cal A} -\langle{\cal A} \rangle)^2 \rangle
=\langle {\cal A}^2\rangle-\langle{\cal A} \rangle^2.
\end{eqnarray*}
This quantity measures the quantum fluctuation.  We find the dispersion of $y^m$ is 
\begin{eqnarray}
\Delta_{y^m}^2= 
\left[\frac{\Gamma(\frac{5}{3}+\frac{2m}{3})}{\Gamma(\frac{5}{3})}-  \left(\frac{\Gamma(\frac{5}{3}+\frac{m}{3})}{\Gamma(\frac{5}{3})} \right)^2 \right]  
\left[\frac{9(\beta^2+A^2t^2)}{2\beta}\right]^{\frac{2m}{3}}
\label{dis y}. 
\end{eqnarray}
and that of the torsion is 
\begin{eqnarray}
\Delta^{2}_{T}&=&\left(\frac{5}{2}+\eta \frac{A^2t^2}{\beta^2}\right)\frac{1}{\Gamma\left(\frac{5}{3}\right)}
\left[\frac{2\beta}{9(\beta^2+A^2t^2)}\right]^{\frac{2}{3}} 
\label{dis T},\nonumber\\
\eta &=& \frac{9}{2}-4\Gamma\left(\frac{4}{3}\right)^2/\Gamma\left({\frac{5}{3}}\right) \simeq 0.9667.
\end{eqnarray}
Then $\Delta_{y^{m}}$ for $m<0$ becomes small and $\Delta_T$ grows as $|t| \rightarrow \infty$. 
Thus in the large $|t|$ region, quantum effect for the $y^m$/tortion is smaller/larger than that in the $|t|  \sim  0$ region.   

We comment that time dependence of a metric similar to (\ref{exv3}) 
have been found in semi-classical analysis of the two dimensional gravity coupled to a scalar field proposed by Jackiw \cite{Jackiw:1984je},
\begin{equation}
S=\int{d^2x}\sqrt{-g}\phi(R-\Lambda),
\end{equation}
where $R$ and $\Lambda$ are Ricci scalar and cosmological constant respectably, and $\phi$ is the scalar field.
Equations of motion become 
\begin{eqnarray}
&&R-\Lambda=0,\nonumber\\
&&\nabla^2\phi-\Lambda\phi=0.
\end{eqnarray}
Provided ansatz for the metric and the scalar field
\begin{eqnarray}
ds^2&=&-dt^2+a^2(t)dx^2, 
\nonumber\\
\phi&=&\phi(t),
\end{eqnarray} 
where $a(t)$ is assumed to take the form $a^2(t)\sim t^{-\alpha}$ with $\alpha$ constant.
Then we have
\begin{equation}
\ddot{\phi}-\alpha t^{-1}\dot{\phi}+\Lambda \phi=0.
\end{equation}
This equation has the following solution
\begin{eqnarray}
\phi(t)=t^{\frac{\alpha+1}{2}}J_{\pm\frac{\alpha+1}{2}}(\sqrt{\Lambda}t),
\end{eqnarray} 
where $J_{\nu}(z)$ is the Bessel function. For large $t$ the scalar field increases as $\phi\sim t^{\alpha/2}$ when $\alpha$ is positive.
Flat universe corresponds to the case of the $\Lambda=0$, In this case the scalar field behaves $\phi\sim t^{(\alpha+1)}$ for large $t$.
In both cases of zero/non-zero cosmological constant, the scale factor $a(t)$ shrinks but the scalar field $\phi(t)$ explodes as time evolves when $\alpha$ is positive.
Hence the feature of this simple model is somehow similar to that of the minisuperspace model discussed in this section.
The scalar field plays a role of a trigger to the non trivial dynamics in this model. 
This may serve as a classical counterpart for late time behavior of our minisuperspace model, where the scalar field is associated to the degree of freedom of the torsion.  

We would like to discuss a possible interpretation of the time dependent wave function (\ref{sol9}). 
Wave function in quantum mechanics is regarded as a one-particle state in the Fock space of quantum field theory. According to this conventional interpretation we identify the wave function (\ref{sol9}) with a one-universe state in the Fock space of matrix quantum gravity. The Fock space describes multi-universes, and there are all kinds of one-universe states determined by the choices of $C(\epsilon)$ in (\ref{wp1}). 
A time dependent inner product $\langle t' | t \rangle =\int^{\infty}_{0} \Psi(t',y)^*\Psi(t,y)dy$ will give a transition probability such that the universe is observed again after a time interval $t'-t$. 
For the wave function (\ref{sol9}) we easily calculated the transition probability by using (\ref{wp2}) and found $\langle t'| t \rangle \propto(t'-t)^{-5/3}$. 
It suggests that this one-universe state is unstable and after some time it will decay into another one-universe state  with some branching ratio, if we ever had introduced operators for creation and annihilation of universes.

\section{Discussion and summary}
\label{Dis}
\setcounter{equation}{0}
In this paper, we proposed a new interpretation of the BFSS matrix model. 
We applied the HKK interpretation, which is originally used for IKKT
model by the authors of \cite{Hanada:2006gg}, to the BFSS matrix model.  
In our interpretation, the time dependent matrix in the BFSS model 
can be regarded as a covariant derivative, 
and further it is decomposed into geometrical quantities such as vielbein or spin connection
with explicit time dependence.  Therefore,
our Hamiltonian determines the time evolution of space time.  
Using this Hamiltonian, we wrote down the Schr\"odinger equation 
following to the usual recipe for quantum mechanics.   

Several questions for our interpretation of the BFSS model still
remain open for further investigation. 
Treatment of the large $N$ limit is one of them.  This requires certain 
regularization procedure which is left unspecified in our paper. 
Such regularization is important not only for making more sense out of 
the equation (\ref{sc8}), but also for quantitative argument. 
It is well recognized that a derivative operator can only be 
expressed as a matrix of infinite size.  Therefore, as was done in this paper, 
most natural way is to start with infinite $N$ theory, and employ
a regularization scheme later to make matrix size finite.  The heat kernel
regularization is a possible candidate for this purpose.
On the contrary, one can also choose an opposite direction---starting from finite $N$ matrix
and taking large $N$ limit later. 
In this case, we should find a 
 ``finite $N$ version''  of covariant derivative which only coincides with 
 true covariant derivative after taking large $N$ limit.  
We notice that the D0-brane field theory developed in \cite{Yoneya:2007ed}
could be useful for this problem.

Relationship between M-theory and our formalism is another issue which is not
addressed in this paper.  Since BFSS matrix model is conjectured to be M-theory, 
our result of quantum gravity would also be understood within M-theory in eleven 
dimension.  We hope to understand this issue in future.

Another point is a relationship between the Wheeler-DeWitt equation and our Shr\"odinger equation.
In subsection \ref{HKK_to_BFSS}, we have seen that our Shr\"odinger equation
can be regarded as a Hamiltonian constraint by introducing a redundant degree of freedom
which corresponds to rescaling of time coordinate.  However, we can still ask whether
our constraint coincides with the Wheeler-DeWitt equation derived from 
a $d+1$ dimensional theory with manifest covariance in whole $d+1$ dimensions.
Since the BFSS model is obviously non-relativistic, we don't expect 
a direct connection to a covariant theory, at least straightforwardly.
Another wishful thinking would be to consider a canonical formalism for the $d+1$ dimensional IKKT model after using the HKK interpretation, 
which hopefully results to a large $N$ effective theory with $d+1$ dimensional covariance.  
For the moment, we leave this issue as an open question. 

We have also investigated a toy model of universe in detail via a two dimensional minisuperspace model.  We have observed that
our wave function is completely regular and normalizable even though 
the classical metric have a singularity at $y \sim \infty$. 
Such phenomena of singularity resolution is rather typical in quantum gravity.
There exist at least two notions of singularity: the singularity appears in the minisuperspace metric and that appears in the Schr\"{o}dinger equation of spacetime (the Wheeler-DeWitt equation).  The singularity which appears in the metric corresponds to either curvature singularity or torsion singularity\footnote{To be specific, the curvature singularity happens at $y \to \infty$ if we consider torsionless gravity, however it is possible to have vanishing curvature with divergent torsion if torsion is allowed.} and Einstein's classical relativity breaks down, while the singularity which appears in the Schr\"{o}diner equation is nonessential but simply a regular singular point in an ODE.  
The regularity can be verified in our wavefunction (\ref{sol9}), and singularity avoidance is achieved via an exponential fall off of the wave function while approaching the point of classical singularity.  Similar scenario of singularity avoidance was also found in the approach of loop quantum gravity\cite{Ashtekar:2006rx}.   
Even if we find no regular solution to the Schr\"{o}diner equation, it is very likely the situation can be improved after including (infinitely) many higher spin fields in the large $N$ matrix, which are ignored through this paper for the simplification of discussion.

More comments are in order.
Firstly, we can study more realistic setup of higher dimensional universe 
with nonzero cosmological constant.  In such case,
additional terms which are absent in two dimensional case
appear in the Hamiltonian.  Then it is interesting to make a comparison with the wave function of four dimensional universe in ordinary Einstein gravity as discussed in \cite{Hartle:1983ai} and hopefully we are able to make contact with the cosmological problem by matrix model \cite{Alvarez:1997fy}.  
Secondly, to make a complete and well defined theory in the HKK interpretation, we include finite/infinite numbers of higher spin fields eventually \cite{Saitou:2006ca}.  
Thirdly, the matrix model in the section \ref{MSSM} is nothing but the $c=1$ matrix model without potential. One expect that there are some hints in this direction as discussed in  \cite{Karczmarek:2003pv}.  Finally, one may extend our study not only to a cosmological spacetime, but also to minisuperspace models with black hole geometry.
We hope to report some results elsewhere.

\section*{Acknowledgements}

The authors would like to thank Kazuyuki Furuuchi, Masanori Hanada, Takayuki Hirayama, Pei-Ming Ho, Hsien-chung Kao, Hikaru Kawai, Feng-Li Lin, Takashi Saitou, Yuuichirou Shibusa, Fumihiko Sugino, Asato Tsuchiya, Tamiaki Yoneya, 
and members of string focus group in Taiwan for useful discussions. 
T.M would like to thank the hospitality of the Okayama Institute for Quantum Physics. 
WYW would like to thank the hospitality of the Leung Center for Cosmology and Particle Astrophysics.
S. Z. is grateful to organizers and participants of KEK Theory Workshop 2008 and members of theory group at RIKEN for valuable discussions, comments and their hospitality. 
This work is supported in part by Taiwan NSC grant NSC 97-2119-M-002-001, NSC 96-2811-M-002-018.

\appendix

\section{Comment on canonical momentum}
\label{momentum}
In this Appendix we give a formal argument to define the canonical conjugate momentum from a Lagrangian which has been given HKK prescription, instead of directly applying HKK prescription to the conjugate momentum, 
To this end  it is instructive to work out the procedure of quantization in some detail. 
As we employ the matrix notation (\ref{Xexpansion}),  the Lagrangian can be written in terms of the adjoint fields as
\bea
L
&=&{1\over 2}h^{AB}\dot{x}_{(i)A}\dot{x}^{(i)}_B
-\frac{1}{4}{f^{AB}}_E{f^{CD}}_F h^{EF} x_{(i)A}x_{(j)B}x^{(i)}_Cx^{(j)}_D,
\eea
where $h^{AB}=t^A_{ab}t^B_{ba}$ and $f^{ABC}$ is the structure constant of $SU(N)$.
We have been employing $A_0=0$ gauge.
The conjugate momentum of $x^{(i)}_A$ is defined as
\begin{eqnarray}
\pi^{(i)A} = \frac{\partial L}{\partial \dot{x}_{(i)A}}=h^{AB}\dot{x}^{(i)}_B,
\label{sc2_2}
\end{eqnarray}
where $\pi^{(i)A}=h^{AB}\pi_B^{(i)}$ and $\pi_A^{(i)}$ is defined through $\Pi^{(i)}_{ab}=\pi_A^{(i)} t^A_{ab}$.
The Legendre transformation gives us the BFSS Hamiltonian
\begin{eqnarray}
H
&=&\frac{1}{2}h_{AB}\pi_{(i)}^{A}\pi^{(i)B}
+\frac{1}{4}{f^{AB}}_E{f^{CD}}_F h^{EF} x_{(i)A}x_{(j)B}x^{(i)}_Cx^{(j)}_D ,
\label{sc3}
\end{eqnarray}
where $h_{AB}$ is an inverse matrix of $h^{AB}$.
The Poisson bracket of $X_{(i)}$ and $\Pi^{(j)}$ is written as
\bea
\left\{x_A^{(i)}, \pi_B^{(j)}\right\}_{PB}=\delta^{(i)(j)}  h_{AB} .
\eea
Upon quantization the Poisson bracket becomes the canonical commutator, and $\pi^{(i)}$ acts as $-i\partial/\partial x_{(i)}$ on the wave function $\Psi(x)$.
The Schr\"odinger equation then becomes
\begin{eqnarray}
i\frac{\partial}{\partial t}\Psi(x)
&=&
\left(-\frac{1}{2} h^{AB}\frac{\partial}{\partial x_{(i)}^A}\frac{\partial}{\partial x^{(i)B}}
+\frac{1}{4}{f^{AB}}_E{f^{CD}}_F h^{EF} x_{(i)A}x_{(j)B}x^{(i)}_Cx^{(j)}_D  \right)\Psi(x) .
\label{sc5'}
\end{eqnarray}

The explicit form of the BFSS Lagrangian after the application of the HKK prescription is 
\begin{eqnarray}
L&=&-\int d^dx \sqrt{h_d}\;tr_{\hat{g}}
\left\{ \frac{1}{2}\dot{\nabla}_i\dot{\nabla}_i
-\frac{1}{4}[\nabla_i, \nabla_j]^2 \right\}
\\
&=&-\int d^dx \sqrt{h_d}\;tr_{\hat{g}}
\left\{ \frac{1}{2}({\dot{e}^K}_{i}  \partial_{K})({\dot{e}^L}_{i} \partial_{L})
+\frac{1}{2}(\dot{\omega_i}^{jk} {\cal O}_{jk}) (\dot{\omega_i}^{lm} {\cal O}_{lm}) 
\right. 
\nonumber \\
&&
\left.
-\frac{1}{4}({T_{ij}}^{K}\partial_{K})({T_{ij}}^{L}\partial_{L})
-\frac{1}{4}({R_{ij}}^{kl}{\cal O}_{kl})({R_{ij}}^{mn}{\cal O}_{mn}) \right\} .
\label{lagrangian}
\end{eqnarray}

Inspired by the formal resemblance of (\ref{kin1}) to (\ref{Xexpansion}), we introduce following notation.
Let  $\tau_A$ represents $\partial_I$ and ${\cal O}_{jk}$, and $E^A_i$ be ${e_i}^I$ and ${\omega_i}^{jk}$, where the index $A$ runs also $I$ and $i,j$. 
Thus (\ref{kin1}) can be written as
\bea
X_{(i)}(t)={E_{(i)}}^A(t) \tau_A ,
\label{formaldecomp}
\eea 
where we have defined ${E_{(i)}}^A={R_{(i)}}^i {E_i}^A$.
However in order to fullfill the analogy, we have to make a crucial assumption that $\tau_A$ and $E^A$ commute with each other.
This is only valid if $E^A$ is not a function of the coordinates $x$ since $\tau_A$ contains the derivative with respect to $x$. 
Only in this limited situation, the analogy would works.
Furthermore, it should be noted that the decomposition (\ref{formaldecomp}) is simply a formal expression and we do not intend that $\tau$ can be expressed as a matrix though its counterpart, those generators of Lie algebra $t^A_{ab}$, can have a matrix representation.  
With these assumptions we also introduce a metric
\bea
{\cal G}_{AB}=\int d^dx \sqrt{h_d}\;tr_{\hat{g}} \tau_A\tau_B,
\eea
and its inverse ${\cal G}^{AB}$, which correspond to $h^{AB}$ and $h_{AB}$ respectively.  
We can rewrite the relevant kinetic term in the Langrangian by using these notations
\bea
L={1\over 2}{\cal G}_{AB} \dot{E}_{(i)}^A \dot{E}_{(i)}^B.
\eea
Then we can compute the canonical momentum 
\bea
\pi_{(i)}^A = {\cal G}^{AB} \pi_{(i)B} = {\cal G}^{AB}{ \delta L \over \delta \dot{E}^B_{(i)}}=\dot{E}^A_{(i)} ,
\eea
which gives (\ref{kin3}).
The Hamiltonian is 
\bea
H&=&\pi_{(i)}^A {\delta L \over \delta \dot{E}_{(i)}^A} - L
\nonumber \\
&=&{1\over 2}{\cal G}_{AB} \pi_{(i)}^A \pi_{(i)}^B
\eea
which gives rise to the kinetic term of (\ref{hal1}).

As we mentioned above this formal argument works only with the assumptions. Without these assumptions it is hard to perform canonical formalism starting from the Lagrangian (\ref{lagrangian}).
We regard that one of the possible origin of these subtleties is due to the large $N$ formalism. 
Thus the process performed in section \ref{HKK-BFSS} to get the canonical momentum (\ref{kin2}) and the Hamiltonian (\ref{hal1}) by applying the HKK prescription directly to the canonical momentum (\ref{sc2}) and the Hamiltonian (\ref{sc3.0}) (which is obtained from the Lagrangian of BFSS model (\ref{BFSSaction})) can be regarded as a finite $N$ regularization.

\section{Embedding BFSS into IKKT}
\label{BFSSintoIKKT}
In this Appendix, we demonstrate that the BFSS model can be embedded into the IKKT model with the HKK interpretation. It will become clear to us which kind of spacetime dynamics are described by the BFSS model with the HKK interpretation. 
Let us examine only the bosonic part of the IKKT action with Lorentzian signature. It is straightforward to include the fermionic sector.  
The IKKT action is
\begin{eqnarray}
S_{IKKT}=\frac{1}{2}Tr [X_{(0)} , X_{(i)}] [X^{(0)} , X^{(i)}]+\frac{1}{4}Tr [X_{(i)},X_{(j)}] [X^{(i)},X^{(j)}]  ,
\label{idn1}
\end{eqnarray}
where the indices $(i),(j)$ run from 1 to $d$. 

We consider the following identification in the large $N$ limit:
\begin{eqnarray} 
X_{(0)}= i{\delta_{(0)}}^{0}D_0, \quad 
Tr=\int dt \;tr ,
\label{idn2}
\end{eqnarray} 
where $D_0=\partial_t+iA_0$ is the gauge covariant derivative and $tr$ is the trace operation of the BFSS model.
This is nothing but the Lorentzian version of the prescription described by \cite{Connes:1997cr}\cite{Taylor:1996ik}.
As a result, we have derived the bosonic part of the BFSS action.

On the other hand, using the HKK interpretation \cite{Hanada:2006ei}, say $X_{(a)}=i{R_{(a)}}^{a}\nabla_a$, the action (\ref{idn1}) becomes 
\begin{eqnarray}
S=\frac{1}{4}Tr[\nabla_a, \nabla_b][\nabla^a,\nabla^b]=
\frac{1}{4}\int dg\int d^{d+1}x\sqrt{h_{d+1}\;}\langle x,g|({R_{ab}}^{cd}{\cal O}_{cd})^2|x,g\rangle.
\label{idn3}
\end{eqnarray}
We have used the relation $Tr F \rightarrow  \int dg\int d^{d+1}x\sqrt{h_{d+1}\;}\langle x,g|F|x,g\rangle$, and ignored torsion here. 
The integral $\int dg \langle g| F |g \rangle$ means the trace operation over the representations of $G$.
We require that applying the HKK interpretation to the BFSS action (\ref{idn2}), say $X_{(i)}=i{R_{(i)}}^{a}\nabla_{a}$, should be consistent with (\ref{idn3}). 
This consistency requires the following restrictions :
\begin{eqnarray}
&&{R_{(0)}}^{0}=1,\quad  {R_{(0)}}^{i}=0,\quad 
{R_{(i)}}^{0}=0, \quad {R_{(i)}}^{j} \in Spin(d), \label{nhkk1}
\\
&& {e_{t}}^{0}=1, \quad {e_{t}}^{i}=0, \quad {e_{I}}^{0}=0, \quad 
{e_{I}}^{i}{e_{J}}^{j}\delta_{ij}=h_d(t,x)_{IJ} ,
\label{nhkk2} 
\\
&& {\omega_{0}}^{jk}=0, \quad {\omega _{j}}^{0k}=0 ,
\label{nhkk3} \\
&&trF=\int d\hat{g}\int d^dx\sqrt{h_d}  \;\langle x,\hat{g}|F|x,\hat{g} \rangle \equiv \int d^dx \sqrt{h_d} \; tr_{\hat{g}} F.\label{nhkk4}
\end{eqnarray}
It is obvious that under these restrictions, we have explicitly broken $G=Spin(d,1)\rightarrow Spin(d)\times {\bf R}$, where $t\in {\bf R}$ and $\hat{g}\in \hat{G}=Spin(d)$.  We claim that (\ref{nhkk1})-(\ref{nhkk4}) are the very HKK interpretation for the BFSS model.   
We remark that the second  condition in (\ref{nhkk3}) comes from the fact that matrix indices $(0)$ and $(i)$ are not mixed with each other in the original BFSS Lagrangian therefore it should also be true for the local Lorentz indices $0$ and $i$ after using the HKK interpretation. 
The conditions (\ref{nhkk2}) and (\ref{nhkk4}) tell us that the HKK interpretation of the BFSS model describes a class of curved spacetime equipped with a metric $ds^2=-dt^2+h_d(t,x)_{IJ}dx^{I}dx^{J}$ $(I,J=1, \cdots,d)$. Then the time parameter $t$ in the BFSS quantum mechanics indeed corresponds to the time coordinate in the curved spacetime.

\section{Torsion, Curvature}
\label{Torsion_Curvature}
In this Appendix, we calculate the commutator $[\nabla_a, \nabla_b]$ explicitly and give definitions for the curvature tensors and the torsion.

Recall that the spin connection is introduced in the covariant derivative such that
\begin{eqnarray}
\nabla_{\mu}V^{a}=\partial_{\mu}V^{a}+\omega_{\mu\;\;b}^{\;\;a}V^{b},
\end{eqnarray} 
where $\mu$ is a curved space index and $a$ is a local Lorentz index.
On the other hand, we have
 \begin{eqnarray}
\nabla_{\mu}V_{a}=\partial_{\mu}V_{a}-\omega_{\mu\;\;a}^{\;\;b}V_{b}
=\partial_{\mu}V_{a}+{\omega_{\mu a}}^{b}V_{b}.
\end{eqnarray} 
To compute the commutator, we first note $\nabla_{a}\nabla_{b}V^{c}$ with local Lorentz indices, that is
\begin{eqnarray}
\nabla_{a}\nabla_{b}V^{c}
&=&
\partial_a(\nabla_bV^c)+ {\omega_{ab}}^{d}(\nabla_dV^c)+{{\omega_{a}}^c}_d(\nabla_bV^d)
\nonumber\\
&=&
\partial_a(\partial_bV^c)
+\omega_{a\;\;d}^{\;\;c}(\partial_b V^d)
+{{\omega_b}^c}_d(\partial_a V^d)
+{\omega_{ab}}^{d}(\nabla_dV^c)
+(\partial_a\omega_{b\;\;d}^{\;\;c})V^d
+{{\omega_a}^c}_d{{\omega_b}^d}_eV^e.
\nonumber \\
\end{eqnarray}
Then the commutator takes following form
\begin{eqnarray}
[\nabla_a, \nabla_b]V^{c}={T_{ab}}^\mu(\nabla_{\mu}V^c)+{{{\cal R}_{ab}}^c}_dV^{d}
\label{com1}
\end{eqnarray}
with 
\begin{eqnarray}
{T_{ab}}^\mu&=&\partial_a {e_{b}}^{\mu}-\partial_b {e_{a}}^{\mu}+{\omega_{ab}}^\mu-{\omega_{ba}}^\mu, 
\\
{{{\cal R}_{ab}}^c}_d&=& e_{a}^{\;\mu}e_{b}^{\;\nu}\left(
 \partial_{\mu}\omega_{\nu\;\;d}^{\;\;c}-\partial_{\nu}\omega_{\mu\;\;d}^{\;\;c}
+\omega_{\mu\;\;e}^{\;\;c}\omega_{\nu\;\;d}^{\;\;e}
-\omega_{\nu\;\;e}^{\;\;c}\omega_{\mu\;\;d}^{\;\;e} \right).
\end{eqnarray}
Here ${T_{ab}}^\mu$ gives rise to torsion and  ${{{\cal R}_{ab}}^c}_d$ is Riemann curvature tensor.
The torsion vanishes if the vielbein and the spin connection satisfy the torsion free condition.
In that case they are no longer independent of each other.
It is convenient to write (\ref{com1}) in another way: 
\begin{eqnarray}
[\nabla_a, \nabla_b]V^{c}&=&{T_{ab}}^\mu(\nabla_{\mu}V^c)+{{{\cal R}_{ab}}^c}_dV^{d}
\nonumber \\
&=&\left(\partial_a e_{b}^{\;\mu}-\partial_b e_{a}^{\;\mu}
+{\omega_{ab}}^{\mu}-{\omega_{ba}}^{\mu} \right)(\partial_{\mu}V^c+\omega_{\mu\;\;d}^{\;\; c}V^{d})
\nonumber\\
&&+e_{a}^{\;\mu}e_{b}^{\;\nu}\left( 
\partial_{\mu}\omega_{\nu\;\;d}^{\;\;c}-\partial_{\nu}\omega_{\mu\;\;d}^{\;\;c}
+\omega_{\mu\;\;e}^{\;\;c}\omega_{\nu\;\;d}^{\;\;e}
-\omega_{\nu\;\;e}^{\;\;c}\omega_{\mu\;\;d}^{\;\;e} 
\right)V^d
\nonumber\\
&=&{T_{ab}}^{\mu}\partial_{\mu}V^c+{{{R}_{ab}}^c}_dV^d,
\end{eqnarray}
where we have defined
\begin{eqnarray}
R_{ab\;\;d}^{\quad c}&=& 
\partial_{a}\omega_{b\;\;d}^{\;\;c}-\partial_{b}\omega_{a\;\;d}^{\;\;c}
+\omega_{a\;\;e}^{\;\;c}\omega_{b\;\;d}^{\;\;e}
-\omega_{b\;\;e}^{\;\;c}\omega_{a\;\;d}^{\;\;e} 
+(\omega_{a\;\;b}^{\;\;e}-\omega_{b\;\;a}^{\;\;e})\omega_{e\;\;d}^{\;\;c}
\nonumber \\
&=&{\cal R}_{ab\;\;d}^{\quad c}+{T_{ab}}^\mu\omega_{\mu\;\;d}^{\;\;c} .
\label{curv}
\end{eqnarray}

\section{Equations of motion}
\label{EOMappendix}
In this Appendix we show the derivation of matrix equation of motion in HKK interpretation.
We first recall
\begin{eqnarray}
[\nabla_{a}, [\nabla_b, \nabla_c]] 
&=& [\nabla_a, ({T_{bc}}^{e}\nabla_{e}+{{\cal R}_{bc}}^{ef}{\cal O}_{ef})]\nonumber\\
&=& (\nabla_a{T_{bc}}^{e})\nabla_{e}+{T_{bc}}^{e}[\nabla_a, \nabla_{e}]
+(\nabla_a{{\cal R}_{bc}}^{ef}){\cal O}_{ef}
+{{\cal R}_{bc}}^{ef}[\nabla_a, {\cal O}_{ef}]
\nonumber\\
&=&(\nabla_a{T_{bc}}^{e})\nabla_{e}+{T_{bc}}^{e} ({T_{ae}}^{\nu}\nabla_{\nu}
+{{\cal R}_{ae}}^{df}{\cal O}_{df})
\nonumber \\
&&+(\nabla_a{{\cal R}_{bc}}^{ef}){\cal O}_{ef}
+{{\cal R}_{bc}}^{ef}\frac{1}{2}(\eta_{ae}\nabla_f-\eta_{af}\nabla_e)
\nonumber\\
&=&[(\nabla_a{T_{bc}}^{e})+{T_{bc}}^{d} {T_{ad}}^{e}+{\cal R}_{bca}^{\quad e}]\nabla_{e}
+[{T_{bc}}^{e}{{\cal R}_{ae}}^{df}+\nabla_a{{\cal R}_{bc}}^{df} ]{\cal O}_{df} .
\label{dcom1}
\end{eqnarray}

The equation of motion in the BFSS model is
\begin{eqnarray}
[D^0, [D_0, X_{(i)}]] + [X^{(j)}, [X_{(j)},X_{(i)}]]&=&0, \label{apeom1}\\
{}[X^{(i)},[X_{(i)}, D_0]]&=&0. \label{apeom2}
\end{eqnarray}
Second equation gives Gauss low constraint.
Let us consider (\ref{apeom1}) first.  After we take $D_0=\partial_t$ gauge,  
applying HKK identification, (\ref{apeom1}) becomes
\begin{eqnarray}
\ddot{\nabla}_{i}+ [\nabla^{j}, [\nabla_{j},\nabla_{i}]]=0,
\label{apeom3}
\end{eqnarray}
where $\ddot{\nabla}$ means $\frac{\partial^2}{\partial t^2}\nabla_i=
\frac{\partial^2}{\partial t^2}{e}_{i}^{\;\;I}\partial_{I}+\frac{\partial^2}{\partial t^2} {\omega}_{i}^{\;\;jk}{\cal O}_{jk}$.
After using (\ref{dcom1}), we obtain
\begin{eqnarray}
\ddot{e}_{i}^{\;\;I}\partial_{I}+\ddot{\omega}_{i}^{\;jk}{\cal O}_{jk}
+\nabla^k{T_{ki}}^{j}+\eta^{lk}{T_{ki}}^{m} {T_{lm}}+{ {\cal R}_{i}}^{j} \nabla_{j}
+[{T_{mi}}^{n}{\cal R}^{m}_{\;nkl}+\nabla^{j}{\cal R}_{jikl}]{\cal O}^{kl}=0.
\label{apeom4}
\end{eqnarray}
We also defined the $d$-dimensional Ricci tensor as $ {\cal R}_{i}^{\;j} =\delta^{kl} { {\cal R}_{kil}}^{j}$.  
From (\ref{apeom4}) we obtain
\begin{eqnarray}
&&\Big( \ddot{e}_{i}^{\;\;I}
+(\nabla^k{T_{ki}}^{j})e_{j}^{\;\;I}+\eta^{lk}{T_{ki}}^{m} {T_{lm}}^{I}+{ {\cal R}_{i}}^{I} \Big)\partial_{I}
\nonumber\\
&&+\Big( \ddot{\omega}_{ikl}
+[\nabla^k{T_{ki}}^{j}+\eta^{lk}{T_{ki}}^{m} {T_{lm}}^{j}+{ {\cal R}_{i}}^{j}]\omega_{jkl}
+{T_{mi}}^{n}{\cal R}^{m}_{\;nkl}+\nabla^{j}{\cal R}_{jikl} \Big){\cal O}^{kl}=0. 
\label{apeom5}
\end{eqnarray}
Here we assume that $\partial_{I}$ and ${\cal O}_{kl}$ form a part of independent basis of infinite matrix. 
Thus the equation of motion breaks into two independent part:
\begin{eqnarray}
 \ddot{e}_{i}^{\;\;I}
+(\nabla^k{T_{ki}}^{j})e_{j}^{\;\;I}+\eta^{lk}{T_{ki}}^{m} {T_{lm}}^{I}+{ {\cal R}_{i}}^{I}&=&0, \label{apeom6}\\
 \ddot{\omega}_{ikl}
-\ddot{e}_{i}^{\;\;I}e_{I}^{\;\;j}\omega_{jkl}
+{T_{mi}}^{n}{\cal R}^{m}_{\;nkl}+\nabla^{j}{\cal R}_{jikl}&=&0.  \label{apeom7}
\end{eqnarray}
For completeness, we rewrite (\ref{apeom6}) and (\ref{apeom7}) in terms of $R$ tensor defined  in (\ref{curv})
\begin{eqnarray}
\ddot{e}_{i}^{\;\;I}
+(\nabla^k{T_{ki}}^{j})e_{j}^{\;\;I}+\eta^{lk}{T_{ki}}^{m} {T_{lm}}^{I}
-{T_{ji}}^{k}\omega^{k}_{\;jI}+{R_{i}}^{I} &=&0, \\
 \ddot{\omega}_{ikl}
-\ddot{e}_{i}^{\;\;I}e_{I}^{\;\;j}\omega_{jkl}
+{T_{mi}}^{n}(R^{m}_{\;nkl}-\eta^{mp}{T_{pm}}^{j}\omega_{jkl})+\nabla^{j}(R_{jikl}-{T_{ij}}^{m}\omega_{mkl})&=&0.
\label{apeom8}
\end{eqnarray}
Similarly, (\ref{apeom2}) becomes
\begin{eqnarray}
&& \left(
e_{i}^{\;I}\partial_{I} \dot{e}_{i}^{\;J}-\dot{e}_{i}^{\;I}\partial_{I} e_{i}^{\;J}
+\omega_{ii}^{\;\; k}\dot{e}_{k}^{\;J}-\dot{\omega}_{ii}^{\;\;k}e_{k}^{\;J} 
\right)\partial_{J}
\nonumber\\
&&+\left( 
\partial_i\dot{\omega}_{i}^{\;kl}-\dot{\partial}_{i}\omega_{i}^{\;kl}
+\omega_{i\;\;j}^{\;\;k}\dot{\omega}_{i}^{\;jl}
-\dot{\omega}_{i\;\;j}^{\;\;k}\omega_{i}^{\;jl}
+\omega_{ii}^{\;\; j}\dot{\omega}_{j}^{\;kl}
-\dot{\omega}_{ii}^{\;\; j}\omega_{j}^{\;kl}
\right){\cal O}_{kl}=0
\label{apeom9}
\end{eqnarray}
where $\dot{\partial}_{i} \omega_{i}^{\;\;kl} =\dot{e}_{i}^{\;I}\partial_{I} \omega_{i}^{\;\;kl} $. 
Finally, we have equations of motion:
\begin{eqnarray}
e_{i}^{\;I}\partial_{I} \dot{e}_{i}^{\;J}-\dot{e}_{i}^{\;I}\partial_{I} e_{i}^{\;J}
+\omega_{ii}^{\;\; k}\dot{e}_{k}^{\;J}-\dot{\omega}_{ii}^{\;\;k}e_{k}^{\;J} &=&0,
\label{apeom10} \\
\partial_i\dot{\omega}_{i}^{\;kl}-\dot{\partial}_{i}\omega_{i}^{\;kl}
+\omega_{i\;\;j}^{\;\;k}\dot{\omega}_{i}^{\;\;jl}
-\dot{\omega}_{i\;\;j}^{\;\;k}\omega_{i}^{\;\;jl}
+\omega_{ii}^{\;\; j}\dot{\omega}_{j}^{\;kl}
-\dot{\omega}_{ii}^{\;\;j}\omega_{j}^{\;kl}
&=&0. \label{apeom11}
\end{eqnarray}

\section{Proof of the orthogonality (\ref{sol2})}
\label{orthogonality}
We give a proof of the orthogonality (\ref{sol2}). 
The left hand side of (\ref{sol2}) can be written as
\bea
{1\over 2}\lim_{a \to 0}\left[
\int_0^\infty dx x e^{-a^2 x^2}J_{\pm{2\over 3}}(\sqrt{\epsilon}x)
J_{\pm{2\over 3}}(\sqrt{\epsilon'}x)
\right],
\eea
where we have multiplied the identity $e^{-a^2 x^2}|_{a=0}=1$.
Now we apply the formula
\begin{eqnarray}
\int_0^\infty dx\;e^{-a^2x^2} xJ_{\nu}\left(px\right)J_{\nu}\left(qx\right)=
\frac{1}{2a^2}e^{-\frac{p^2+q^2}{4a^2}}I_{\nu}\left(\frac{pq}{2a^2}\right) ,
\label{sol3}
\end{eqnarray}
that is valid when $Re\nu>-1$ and $|Arg \;a|<\frac{\pi}{4}$. 
Here $I_{\nu}(z)$ is the modified Bessel function. 
We have
\bea
{1\over 2}\lim_{a \to 0}\left[
{1\over 2a^2}e^{-{\epsilon+\epsilon' \over 4a^2}}
I_{\pm {2\over 3}}\left({\sqrt{\epsilon\epsilon'}\over 2a^2}\right)
\right].
\label{left_hand_side_3.7}
\eea
Using asymptotic formula for $I_{\nu}(z)$ at large $z$,
\begin{eqnarray}
I_{\nu}(z) \sim \frac{e^z}{\sqrt{2\pi z}} ,
\label{sol4}
\end{eqnarray} 
(\ref{left_hand_side_3.7}) becomes
\bea
\lim_{a \to 0}\left[
{1\over 4a\sqrt{\pi \sqrt{\epsilon\epsilon'}}}e^{-{(\sqrt{\epsilon}-\sqrt{\epsilon'})^2 \over 4a^2}}
\right]
=\delta(\epsilon-\epsilon'),
\eea
where we have used $\delta(\sqrt{\epsilon}-\sqrt{\epsilon'})=2\sqrt{\epsilon}\delta(\epsilon-\epsilon')$.


 \end{document}